\documentclass{article}
\usepackage{arxiv}
\usepackage{amsmath}
\usepackage[utf8]{inputenc} 
\usepackage[T1]{fontenc}  
\usepackage{url}            
\usepackage{booktabs}       
\usepackage{amsfonts}       
\usepackage{nicefrac}       
\usepackage{microtype}      
\usepackage{lipsum}
\usepackage{graphicx}
\usepackage{float}
\usepackage{multirow}
\usepackage{tabularx}
\usepackage{setspace}
\usepackage{enumerate}
\usepackage{makecell}
\usepackage{placeins}
\usepackage[colorlinks,allcolors=black]{hyperref}
\graphicspath{ {./images/} }
\DeclareMathOperator*{\argmax}{arg\,max}

\usepackage{hyperref}       
\usepackage{amsthm}
\newtheorem{Theorem}{Theorem}
\newtheorem{Lemma}[Theorem]{Lemma}
\graphicspath{ {./images/} }

\title{A Generalized Family of Exponentiated Composite Distributions}

\author{
Bowen Liu \\
 Department of Mathematical Sciences\\
    University of Nevada, Las Vegas\\ 
    NV 89154\\
  \texttt{bowen.liu@unlv.edu} \\
   \And
 Malwane A. Ananda\\
 Department of Mathematical Sciences\\
    University of Nevada, Las Vegas\\ 
    NV 89154\\
  \texttt{malwane.ananda@unlv.edu} \\
}

\begin{document}
\maketitle

\begin{abstract}

In this paper, we propose a new class of distributions by exponentiating the random variables associated with the probability density functions of composite distributions. We also derive some mathematical properties of this new class of distributions including the moments and the limited moments. Specifically, two special models in this family are discussed. Two real data sets were chosen to assess the performance of these two special exponentiated composite models. When fitting to these two data sets, theses two special exponentiated composite distributions demonstrate significantly better performance compared to the original composite distributions.
\end{abstract}


\section{Introduction}
The concept of composite distributions is widely used in different aspects such as modeling insurance claim size data \cite{ananda2005,scollnik_composite_2007,Bak15,brazauskas_modeling_2016,grun2019, ig_pareto, liu_analyzing_2022}, fitting survival time data \cite{cooray_weibull_2010} and modeling precipitation data \cite{kim_bias_2019}. Such concept demonstrates impressive performances when the data is characterized with very heavy tail, which, common distributions such as normal or exponential distributions cannot capture all the data features. Due to the simplicity and the applicability of the concept, the researchers developed a considerable number of composite distributions including Lognormal-Pareto \cite{ananda2005}, Weibull-Pareto \cite{cooray_weibullpareto_2009} , Weibull-Inverse Weibull \cite{cooray_weibull_2010}, and so on.  

\par The composite distributions seems proper when modeling the data with heavy tails. For example, both of the one-parameter Inverse Gamma- Pareto (IG-Pareto) model \cite{ig_pareto} and the one-parameter exponential-Pareto (exp-Pareto) model \cite{Teodorescu2006} were suggested as possible models for insurance data modeling. However, they still cannot provide a satisfactory performance when fitting to well known insurance data sets such as the Danish Fire Insurance data sets. Thus, it is necessary to improve the model. To improve the performance of of the one-parameter IG-Pareto model, Liu and Ananda \cite{liu_analyzing_2022} proposed an exponentiated IG-Pareto model by exponentiating the random variable associated with the probability density function (pdf) of an Inverse Gamma-Pareto distribution. With different data sets, The newly proposed model demonstrated significant improvement from the original model. 

\par In fact, we noted the idea of the exponentiated IG-Pareto model can be generalized to all the existing composite distributions. Thus, in this paper, we propose a generalized family of exponentiated composite distributions with the derivation of some mathematical properties of this family. 

\par The rest of the paper is organized as follows. Section 2 provides the formulation of the generalized exponentiated composite distributions as well as some mathematical properties of this new family of distributions. We then briefly discussed two special exponentiated composite distributions (exponentiated IG-Pareto and exponentiated exp-Pareto) in Section 3. In Section 4, we introduce a parameter estimation method for exponentiated composite models. The simulation results for the estimation method is also provided in Section 4. Two numerical examples are presented in Section 5. To conclude, a discussion is provided in Section 6. 

\section{A Generalized Class of Exponentiated Composite Distributions}
\subsection{Model Formulation}
Suppose X is a random variable that takes on non-negative real numbers. Let $f(x)$ be the probability density function (pdf) of $X$. A composite distribution function $f(x)$ is defined 
\cite{scollnik_composite_2007} as follows:
\begin{align}
 f_X(x|\alpha_1,\alpha_2,\theta) = \begin{cases} 
      cf_1(x|\alpha_1,\theta) & x \in [0,\theta) \\
      cf_2(x|\alpha_2,\theta) & x \in [\theta,\infty)
   \end{cases},
\end{align}
where c represents the normalizing constant, $f_1$ is the probability density function of random variable $X$ when $X$ is between $0$ and $\theta$; $f_2$ is the probability density function of the random variable $X$ when $X$ is greater than $\theta$. In real practice, we assume both $f_1$ and $f_2$ are smooth functions on their supports. However, the definition in (1) does not guarantee that $f_X(x)$ is a continuous differentiable function. To define continuous differentiable pdf for composite distributions, the continuity and differentiability conditions are taken into account as follows:
\[\begin{cases}
\text{lim}_{x \rightarrow \theta^-}f_X(x|\alpha_1, \alpha_2,\theta)= \text{lim}_{x \rightarrow \theta^+}f_X(x|\alpha_1, \alpha_2,\theta) \\
\text{lim}_{x \rightarrow \theta^-}\frac{df_X(x|\alpha_1,\alpha_2,\theta)}{dx}= \text{lim}_{x \rightarrow \theta^+}\frac{df_X(x|\alpha_1,\alpha_2,\theta)}{dx}.

\end{cases}
\]
Essentially, the conditions can be summarized in a simpler way as: $f_1(\theta) = f_2(\theta)$ and $f_1^{'}(\theta) = f_2^{'}(\theta)$. 
\par 
\par We apply a power transformation to the random variable $X$ by letting $Y = g(X) = X^{1/\eta}$, where $g$ is monotone increasing for any $\eta>0$. The inverse function $g^{-1}(Y) = Y^{\eta}$ has continuous derivative on $(0,\infty)$ for any $\eta>0$. Then the pdf of $Y$ is given by: 
\begin{align} 
        f_Y(y|\alpha_1,\alpha_2,\theta,\eta) = \begin{cases} 
      cf_1(y^{\eta}|\alpha_1,\theta) \eta y^{\eta-1} & y \in [0,\theta^{\frac{1}{\eta}}) \\
      cf_2(y^{\eta}|\alpha_2,\theta)\eta y^{\eta-1} & y \in [\theta^{\frac{1}{\eta}},\infty)
   \end{cases}
\end{align} 

We shall prove that the pdf of $Y$ is still a continuous differentiable pdf of a composite distribution.
\begin{Theorem}
If $X$ is a random variable associated with a continuous differentiable pdf of a composite distribution, then Y = $X^{\frac{1}{\eta}}$ also has a continuous differentiable pdf of a composite distribution for all $\eta>0$.
\end{Theorem}

\begin{proof}[Proof of Theorem 1]
Let $u = \theta^{\frac{1}{\eta}}$. \\
We first show that $\text{lim}_{y \rightarrow {u}^-}f_Y(y|\alpha_1,\theta,\eta)= \text{lim}_{y \rightarrow u^+}f_Y(y|\alpha_2,\theta,\eta)$.
\\
From (2), we have \begin{align*}
\text{lim}_{y \rightarrow {u}^-}f_Y(y|\alpha_1,\theta,\eta) & =  cf_1(u^{\eta}|\alpha_1,\theta) \eta u^{\eta-1}
\\
& = cf_1(\theta|\alpha_1,\theta) \eta u^{\eta-1}\\
& = cf_2(\theta|\alpha_1,\theta) \eta u^{\eta-1} \\
& = cf_2(u^{\eta}|\alpha_1,\theta) \eta u^{\eta-1} \\
& = \text{lim}_{y \rightarrow {u}^+}f_Y(y|\alpha_2,\theta,\eta)
\end{align*}
\\ Then, we want to show that $\text{lim}_{y \rightarrow u^-}\frac{df_Y(y|\alpha_1,\theta,\eta)}{dy}= \text{lim}_{y \rightarrow u^+}\frac{df_Y(y|\alpha_1,\theta,\eta)}{dy}$.\\ 
By the chain rules, we have: \\
\begin{align*}
\text{lim}_{y \rightarrow u^-}\frac{df_Y(y|\alpha_1,\theta,\eta)}{dy} & = c(f_1 ^{'} (u^\eta)\eta u^{\eta-1}+f_1(u^\eta)(\eta-1)u^{\eta-2})
\\ & = c(f_2 ^{'} (u^\eta)\eta u^{\eta-1}+f_2(u^\eta)(\eta-1)u^{\eta-2}) \\
&  = \text{lim}_{y \rightarrow u^+}\frac{df_Y(y|\alpha_2,\theta,\eta)}{dy}
\end{align*}
Thus, $f_Y(y|\alpha_1,\theta,\eta)$ is a continuous differentiable pdf of a composite distribution. 
\end{proof}
Since $f_Y(y)$ is still a composite distribution, we hereby name $f_Y(y)$ as \textbf{the exponentiated composite distribution induced by the parent composite distribution $f_X(x)$}. Correspondingly, we also name $Y$ as \textbf{the exponentiated composite random variable induced by the parent composite random variable} $X$. 

\begin{Lemma}
If $Y$ is a random variable associated with a pdf of an exponentiated composite distribution induced by $X$, where $X$ is associated with a pdf of a continuous differentiable composite distribution, then any exponetiated composite distribution induced by $Y$ is induced by $X$. 
\end{Lemma}

\begin{proof}[Proof of Lemma 1]
The proof is simple. Let $U = Y^{\frac{1}{\gamma}}$\\ Since $Y$ is induced by $X$, there exists $\eta>0$ such that $Y = X^{\frac{1}{\eta}}$. 
\\
Hence,$ U = Y^{\frac{1}{\gamma}} = {(X^{\frac{1}{\eta}})}^{\frac{1}{\gamma}} = X^{\frac{1}{\eta \gamma}}$
Therefore, $U$ is induced by $X$.
\end{proof}
Notice Lemma 1 essentially gives us the general idea that if a random variable $Y$ is induced by $X$, where $X$ is a random variable associated with a composite distribution, then exponentiating $Y$ eventually leads to a distribution of the same type as $Y$. For example, a Weibull-Inverse Weibull composite \cite{cooray_weibull_2010} can be seen as a distribution induced by an Exponential-Inverse Exponential composite. Hence, by exponentiating a Weibull-Inverse Weibull composite random variable, we still get a Weibull-Inverse Weibull composite random variable.  


\subsection{Mathematical Properties}
In this subsection we derive the moments and the limited moments of $Y$ as an induced exponentiated composite random variable of the composite random variable $X$. Before any derivations, we need to guarantee that $Y$ has finite moments if $X$ has finite moments. We first start by proving that $Y$ is $L_p$ integrable as long as $X$ is $L_p$ integrable. 
\begin{Lemma}
Suppose $X$ is a composite random variable. If $X$ is $L_p$ integrable, then the exponentiated composite random variable $Y$ induced by $X$ is $L_p$ integrable. 
\end{Lemma}
\begin{proof}[Proof of Lemma 2]
Suppose $X$ is $L_p$ integrable. Then $E|X|^p < \infty$ for all $p>0$. \\
Based on the definition in (1), $X$ is a nonnegative random variable since both $f_1$ and $f_2$ are valid pdf. Therefore, $E|X|^p = E(X^p) < \infty$ for all $p>0$. \\ 
Since $Y$ is induced by $X$, $Y = X^{\frac{1}{\eta}}$. Then $E(X^p) = E(Y^{\frac{p}{\eta}}) < \infty$. \\
Let $t = \frac{p}{\eta}$. Since $\eta>0$, we have $E(Y^t)<\infty$ for all $t>0$. \\
Thus, $Y$ is $L_p$ integrable if $X$ is $L_p$ integrable. 
\end{proof}
The following derivations are based on the assumption that $X$ is $L_p$ integrable. Assume the $k$-th moment of $X$ is $\mu_k$. 
\subsubsection{$k$-th Moment}
Since we defined that $Y = X^{\eta}$, the $k$-th moment of $Y$ can be simply derived as follows:
\begin{equation}
    E(Y^k) = E(X^{\frac{k}{\eta}}) = \mu_{\frac{k}{\eta}}
\end{equation}
Essentially, the moment of $Y$ is a real-valued fractional moment of $X$. 
\subsubsection{Limited $k$-th Moment}
\par Given $b>0$ and a pdf $f_X(x)$ associated with a random variable $X$, the limited loss random variable $X \wedge b$ is defined as following:

\[ X \wedge b =  \begin{cases} 
X & x \in (0,b) \\
b & x \in [b,\infty) \\
\end{cases}
\]

\par Respectively, the limited $t^{th}$ moment of a random variable $X$, denoted by $E[(X\wedge b)^t]$ is defined as: 
\begin{align*}
	E[(X\wedge b)^t] = \int_{0}^{b} x^t f_X(x) dx +\int_{b}^{\infty} b^t f_X(x) dx
\end{align*}
It follows that if $X$ is $L_p$ integrable, then the $t$-th limited moment of $X$ exists. \par For any valid probability distribution defined on $\mathcal{R}$, the Cumulative Distribution Function (CDF) must exist. Thus, without loss of generality, let the CDF of $X$ be $F_X(x)$ and define the incomplete moment function as $M(u; r) = \int_{0}^{u} x^r f_X(x) dx$. Then the $t$-th limited moment of $X$ can be rewritten as follows:
\begin{align*}
    	E[(X\wedge b)^t] = M(b; t) +b^t (1-F_X(b))
\end{align*}
Suppose $Y = X^{\frac{1}{\eta}}$. Then the corresponding pdf of $Y$ is $f_Y(y) = f_X(y^{\eta}) \eta y^{\eta-1}$. Therefore, we can explicitly represent the $t$-th limited moment of $Y$ as follows, with the incomplete moment function of $X$ and the CDF of $X$:
\begin{equation}
\begin{aligned}
E[(Y\wedge b)^t] & = \int_{0}^{b} y^t f_Y(y) dy +\int_{b}^{\infty} b^t f_Y(y) dy \\
& = \int_{0}^{b^{\eta}} y^t f_X(y^{\eta})\eta y^{\eta-1} dy +\int_{b}^{\infty} b^t f_X(y^{\eta}) \eta y^{\eta-1} dy \\
& = \int_{0}^{b^{\eta}} x^{\frac{t}{\eta}} f_X(x) dx +\int_{b^{\eta}}^{\infty} b^t f_X(x)  dx \\
& = M(b^{\eta}; \frac{t}{\eta}) + b^t (1-F_X(b^\eta))
\end{aligned}
\end{equation}
Moreover, assume $X$ is a random variable associated with a continuous differentiable pdf with a form in (1).  Define the the following:
\begin{align*}
    M_1 (u; r) & =\int_{0}^{u} x^r f_1(x) dx  \\
    M_2 (u; r) & = \int_{0}^{u} x^r f_2(x) dx \\ 
    F_1(u) & = \int_{0}^{u} f_1(x) dx \\
    F_2(u) & = \int_{0}^{u} f_2(x) dx \\
\end{align*}
Then $E[(X\wedge b)^t]$ can be expressed explicitly with the above quantities as follows:
	\[ E[(X\wedge b)^t] =  \begin{cases} 
	 c M_1 (b; t) +cb^t (F_1(\theta) - F_1(b)) +cb^t (1-F_2(\theta))& b \in (0,\theta) \\
	 c M_1(b;t) +c b^t (1-F_2(b))& b = \theta \\
	 c M_1(\theta; t)+ cM_2(b;t) -cM_2(\theta; t)+cb^t (1-F_2(b))& b \in (\theta, \infty), \\
	\end{cases}
	\]
Suppose $Y$ is an exponentiated composite random variable induced by $X$, where the general form of $Y$ is defined in (2). We can also explicitly represent $E[(Y \wedge b)^t]$ as follows:
\begin{equation}
	 E[(Y\wedge b)^t] =  \begin{cases} 
	 c M_1 (b^{\eta}; \frac{t}{\eta}) +cb^t (F_1(\theta) - F_1(b^{\eta})) +cb^t (1-F_2(\theta))& b \in (0,\theta^{\frac{1}{\eta}}) \\
	 c M_1 (b^{\eta}; \frac{t}{\eta})  +cb^t (1-F_2(\theta))&  b = \theta^{\frac{1}{\eta}} \\
	 c M_1(\theta; \frac{t}{\eta})+ cM_2(b^{\eta};\frac{t}{\eta}) -cM_2(\theta; \frac{t}{\eta})+cb^t (1-F_2(b^{\eta}))& b \in (\theta^{\frac{1}{\eta}}, \infty), \\
	\end{cases}
\end{equation}
\section{Special Distributions}
\subsection{Two-Parameter Exponentiated inverse Gamma-Pareto Model}
\par The one-parameter composite IG-Pareto model was introduced by Aminzadeh and Deng \cite{ig_pareto}. Suppose a random variable $X$ follows a one-parameter composite Inverse Gamma-Pareto distribution such that the pdf of $X$ is as follows:
\begin{align*} f_{X}(x|\theta) = \begin{cases} 
\frac{c(k\theta)^{\alpha}x^{-\alpha-1}e^{-\frac{k\theta}{x}}}{\Gamma(\alpha)} & x \in [0,\theta) \\
\frac{c(\alpha-k)\theta^{\alpha-k}}{x^{\alpha-k+1}} &  x \in [\theta,\infty) ,\\
\end{cases}
\end{align*}where, $c = 0.711384, k = 0.144351, a = 0.163947, \alpha = 0.308298$.
\par By utilizing (2), Liu and Ananda \cite{liu_analyzing_2022} developed the two-parameter exponentiated Inverse Gamma-Pareto model. The model has the pdf as follows:
\begin{equation}
   f_{Y}(y|\theta, \eta) = \begin{cases} 
          \frac{c(k\theta)^{\alpha}(y^{\eta})^{-\alpha-1}e^{-\frac{k\theta}{y^{\eta}}}}{\Gamma(\alpha)}\eta y^{\eta-1} & y \in [0,\theta^{\frac{1}{\eta}}) \\
          \frac{c(\alpha-k)\theta^{\alpha-k}}{(y^{\eta})^{\alpha-k+1}}\eta y^{\eta-1} &  y \in [\theta^{\frac{1}{\eta}},\infty) \\
       \end{cases} 
       \end{equation}
The $t$-th raw moment can be easily derived as follows:
\begin{equation}
    E(Y^t) = c(\frac{(k \theta)^{\frac{t}{\eta}}}{\Gamma(\alpha)} \Gamma(\alpha-\frac{t}{\eta},k) - \frac{(\alpha-k) \theta ^{k-\alpha+\frac{t}{\eta}}}{k-\alpha+\frac{t}{\eta}}),
\end{equation}
where, $\Gamma(.,.)$ stands for an upper incomplete gamma function, $\Gamma(\alpha,x) = \int_{x}^{\infty} t^{\alpha-1}e^{-t}dt $. When $k-\alpha+\frac{t}{\eta}<0$, the $t$-th moment is finite. 
The $t$-th limited moment of $Y$ was derived by Liu and Ananda \cite{liu_analyzing_2022} as follows: 
	\[ E[(Y\wedge b)^t] =  \begin{cases} 
	c[\frac{\Gamma(\alpha-\frac{t}{\eta},\frac{k\theta}{b^{\eta}})(k\theta)^{\frac{t} {\eta}}+b^t \Gamma(\alpha,k)-b^t \Gamma(\alpha,\frac{k\theta}{b^{\eta}})}{\Gamma(\alpha)} + b^t] & b \in (0,\theta^{1/\eta}) \\
	 c[\frac{\Gamma(\alpha-\frac{t}{\eta},k)(k\theta)^{\frac{t}{\eta}}}{\Gamma(\alpha)}+b^t]& b = \theta^{1/\eta} \\
	 c \{\frac{\Gamma(\alpha-\frac{t}{\eta},k)(k\theta)^{\frac{t}{\eta}}}{\Gamma(\alpha)}+ \frac{(\alpha-k)[b^{t-\eta(\alpha-k)}\theta^{\alpha-k}-\theta^{\frac{t}{\eta}}]}{k-\alpha+\frac{t}{\eta}}+b^{t-(\alpha-k)\eta }\theta^{\alpha-k}\}& b \in (\theta^{1/ \eta}, \infty), \\
	\end{cases}
	\]

\par
\begin{figure}
	\centering
	\includegraphics[width=1\linewidth]{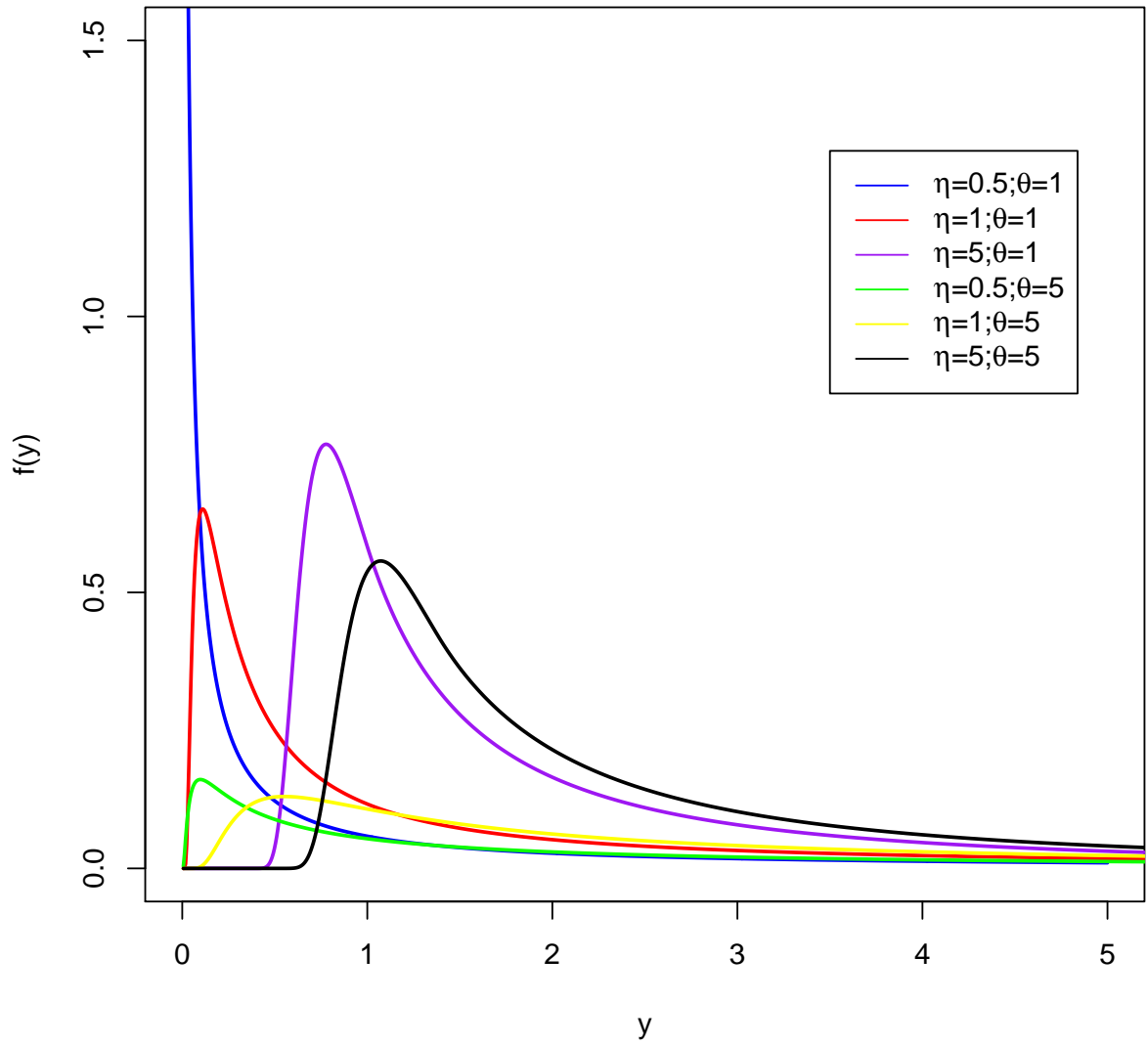}
	\caption{Plots of the exponentiated Inverse Gamma-Pareto density function for some parameter values.}
	\label{fig:exp_ig_pareto}
\end{figure}

 Figure \ref{fig:exp_ig_pareto} is the plot of exponentiated IG-Pareto distributions for different values of $\eta$ and $\theta$. The extra exponent parameter $\eta$ introduces the additional flexibility to the original model. When $\eta \geq 1$, the distribution is associated with a unimodal pdf. 
\subsection{Two-Parameter Exponentiated exp-Pareto Model}
\par The one-parameter composite IG-Pareto model was introduced by Teodorescu \cite{Teodorescu2006}. Suppose a random variable $X$ follows a one-parameter composite exp-Pareto distribution such that the pdf of $X$ is as follows:
\begin{align*} f_{X}(x|\theta) = \begin{cases} 
c (\frac{\alpha+1}{\theta}) e^{-\frac{(\alpha+1) x}{\theta}} &  x \in [0,\theta) \\
c \alpha \frac{\theta^{\alpha}}{x^{\alpha+1}} &  x \in [\theta,\infty) ,\\
\end{cases}
\end{align*}
where, $c = 0.574, \alpha = 0.349976$.
By utilizing (2), the corresponding two-parameter exponentiated exp-Pareto pdf is as follows: 
\begin{equation} f_{Y}(y|\theta, \eta) = \begin{cases} 
c (\frac{\alpha+1}{\theta}) e^{-\frac{(\alpha+1) y^\eta}{\theta}} \eta y^{\eta-1}  &   y \in [0,\theta^{\frac{1}{\eta}}) \\
c \alpha \frac{\theta^{\alpha}}{(y^{\eta})^{\alpha+1}} \eta y^{\eta-1} &   y \in [\theta^{\frac{1}{\eta}},\infty) ,\\
\end{cases}
\end{equation}

	\begin{figure}
	\centering
	\includegraphics[width=1\linewidth]{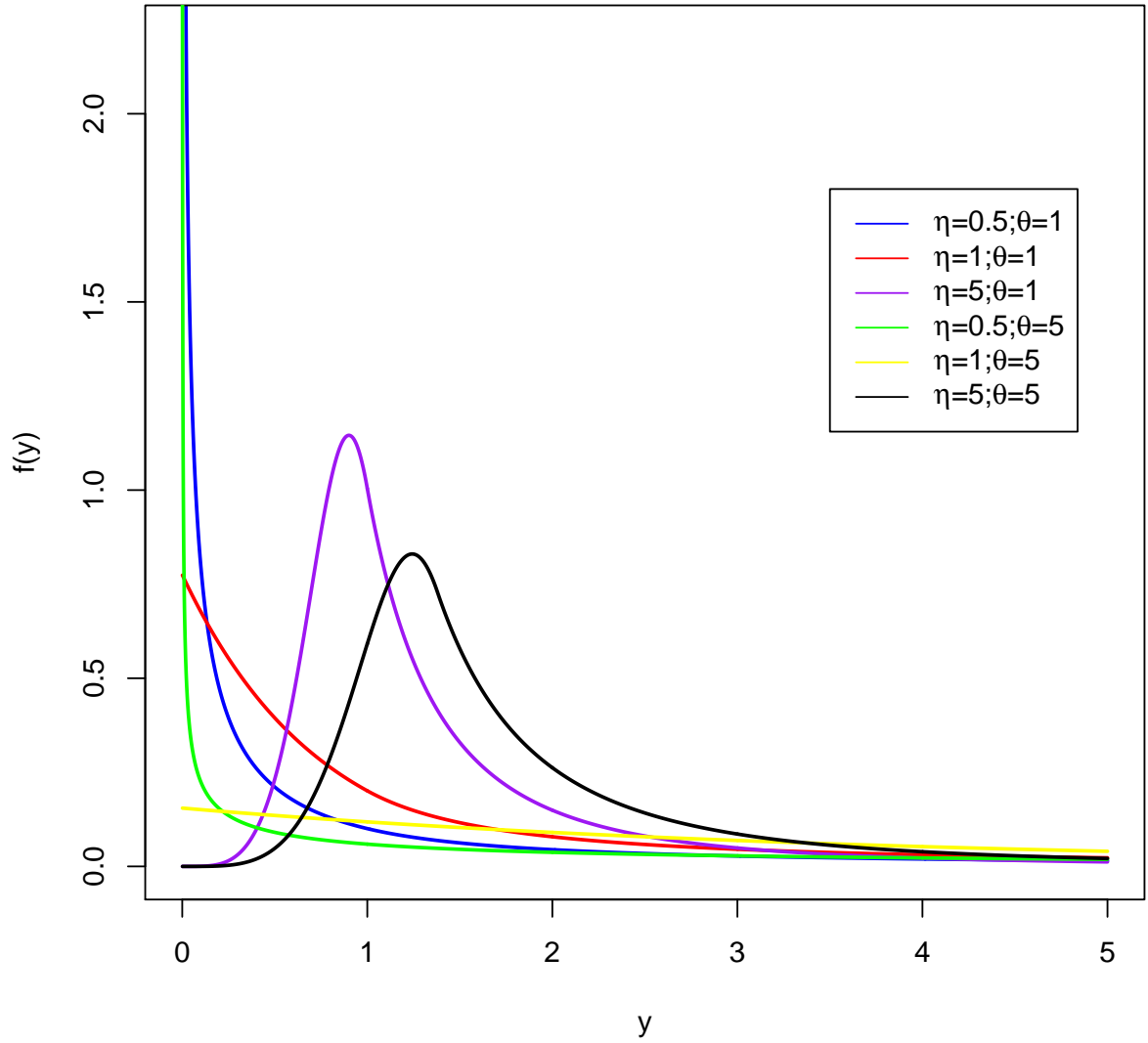}
	\caption{Plots of the exponentiated exp-Pareto density function for some parameter values.}
	\label{fig:eep}
\end{figure}
The $t$-th raw moment of the exponentiated exp-Pareto is derived as follows:
\begin{equation}
    E(Y^t) = 
    c(\frac{\theta}{\alpha+1})^{\frac{t}{\eta}}(\Gamma(\frac{t}{\eta}+1)-\Gamma(\frac{t}{\eta}+1,\alpha+1))+\frac{c \alpha \theta^{\frac{t}{\eta}}}{\alpha-\frac{t}{\eta}},
\end{equation}
where, $ \Gamma(\alpha)$ represents a Gamma function and $\Gamma(\alpha) = \int_{0}^{\infty} t^{\alpha-1}e^{-t}dt $. Here, $\Gamma(.,.)$ stands for an upper incomplete gamma function with $\Gamma(\alpha,x) = \int_{x}^{\infty} t^{\alpha-1}e^{-t}dt $. When $\frac{t}{\eta} < \alpha$, the $t$-th moment is finite. 
The $t$-th limited moment of the exponentiated exp-Pareto is:
\[ E[(Y\wedge b)^t] =  \begin{cases} 
	c\{(\frac{\theta}{\alpha+1})^{\frac{t}{\eta}}[\Gamma(\frac{t}{\eta}+1)-\Gamma(\frac{t}{\eta}+1,\frac{\alpha+1}{\theta}b^\eta)]+b^t (e^{-\frac{\alpha+1}{\theta}b^{\eta}} - e^{-\alpha-1}) + b^t\}  &  b < \theta^{1/\eta} \\
    c\{(\frac{\theta}{\alpha+1})^{\frac{t}{\eta}}[\Gamma(\frac{t}{\eta}+1)-\Gamma(\frac{t}{\eta}+1,\alpha+1)]+b^t \}	& b = \theta^{1/\eta} \\
	c\{(\frac{\theta}{\alpha+1})^{\frac{t}{\eta}}[\Gamma(\frac{t}{\eta}+1)-\Gamma(\frac{t}{\eta}+1,\alpha+1)]+\frac{c\alpha \theta^{\alpha}}{\frac{t}{\eta}-\alpha}(b^{t-\frac{\alpha}{\eta}} - \theta^{\frac{t}{\eta}-\alpha}) +b^t \} &  b > \theta^{1/\eta}, \\
	\end{cases}
	\]
\par Figure \ref{fig:eep} is the plot of exponentiated exp-Pareto distributions for different values of $\eta$ and $\theta$. When $\eta >1$, the pdf is unimodal. For the case that $\eta<1$, the pdf is monotonically decreasing. 

\section{Parameter Estimation and Simulation Studies}
\subsection{Parameter Search Method}
In this section, we want to briefly explain the parameter estimation of the two-parameter exponentiated composite models. Liu and Ananda \cite{liu_analyzing_2022} showed that the parameter estimation of the exponentiated IG-Pareto model can be done using a step-wise grid search procedure on the exponent parameter $\eta$. Suppose $L(\boldsymbol{y}|\theta, \eta)$ be the likelihood of an i.i.d sample from a two-parameter exponentiated composite distribution with a pdf defined in (2). Assume within the sample $\boldsymbol{y}$, let $m$ observations are less than $\theta^{\frac{1}{\eta}}$. Now suppose for every fixed $\eta$, the analytical solution of $\frac{\partial L(\boldsymbol{y}|\theta, \eta)}{\partial \theta} =0$ exists where $\theta$ could be represented as a function of $\eta, m$ and $\boldsymbol{y}$. Denote the function as $\theta(\eta, m, \boldsymbol{y})$.        

\begin{enumerate}[I.]
        \item Arrange the observations in a sample in an increasing order such that $y_{(1)} \leq y_{(2)}\leq ... \leq y_{(n)}$. 
        \item  Define a candidate set $\boldsymbol{\eta}_c$ for the exponent parameter $\eta$. 
        \item For each $\eta \in \boldsymbol{\eta}_c$, we start with $m=1$ and calculate the estimate of $\theta$ based on $\theta(\eta, m, \boldsymbol{y})$. Denote this estimate as $\hat{\theta}_{1,\eta}$. If $y_1^{\eta} \leq  \hat{\theta}_{1,\eta} \leq y_2^{\eta}$, then $m = 1$. If not, go to step IV.
        \item Let $m = 2$. calculate the estimate of $\theta$ based on $\theta(\eta, m, \boldsymbol{y})$. Denote this estimate as $\hat{\theta}_{2,\eta}$. If $y_2^{\eta} \leq  \hat{\theta}_{2,\eta} \leq y_3^{\eta}$, then $m = 2$. The above steps will resume till $m$ is detected. Once $m$ is detected, keep $\hat{\theta}_{m,\eta}$ as the estimate of $\theta$ for the corresponding $\eta$. 
        \item The estimates of $\eta$ is then identified as the $\hat{\theta}$ that maximizes the likelihood function:  
        $ L(\boldsymbol{y}|\theta,\eta)$:
        \begin{equation*}
            \hat{\eta} = \argmax_{\eta \in \boldsymbol{\eta}_c}  L(\boldsymbol{y}|\theta,\eta)
        \end{equation*}
        The corresponding estimate of $\theta$ is then determined as $\theta(\hat{\eta}, m, \boldsymbol{y})$.
    \end{enumerate}

With the above algorithm, the accuracy of the estimates from the two-parameter exponentiated IG-Pareto model was demonstrated with limited simulations. Thus, in the next section, we demonstrate that the parameter estimation of the two-parameter exponentiated exp-Pareto distribution can be done based on the above procedure. 

\subsection{Estimation of the parameters of the exponentiated exp-Pareto model}
To utilize the search procedure in 4.1, the closed-form solution of $\frac{\partial L(\boldsymbol{y}|\theta, \eta)}{\partial \theta} =0$ needs to be obtained. Suppose $y_1, y_2, ..., y_n$ are i.i.d two-parameter exponentiated exp-Pareto random variables with the pdf defined in (7) and assume $y_1 < y_2 < ... < y_n$ without loss of generality.  Assume that there exist an value $m \in \{1,2,...,n-1\}$ such that $y_m^{\eta}<\theta<y_{m+1}^{\eta}$. The likelihood $L(\boldsymbol{y}|\theta, \eta)$ could be written as follows:
\begin{equation*}
        \begin{aligned}
        L(\boldsymbol{y}|\theta, \eta) &= \prod_{i = 1}^{m}   c (\frac{\alpha+1}{\theta}) e^{-\frac{(\alpha+1) y_i^\eta}{\theta}} \eta y_i^{\eta-1} 
        \prod_{j = m+1}^{n}c \alpha \frac{\theta^{\alpha}}{(y_j^{\eta})^{\alpha+1}} \eta y_j^{\eta-1} \\
        &= \frac{c^{n}\eta^{n}(\alpha+1)^{m}\alpha^{n-m}(\prod_{i=1}^{m}y_i)^{\eta-1}}{(\prod_{j=m+1}^{n}y_j)^{\alpha \eta+1}} \theta^{\alpha n -(\alpha+1) m} e^{-\frac{\alpha+1}{\theta} \sum_{i=1}^{m}\frac{1}{y_i^{\eta}}} \\
        &= Q \theta^{\alpha n -(\alpha+1) m} e^{-\frac{\alpha+1}{\theta} \sum_{i=1}^{m}\frac{1}{y_i^{\eta}}},
    \end{aligned}
    \end{equation*}
    where
    \begin{equation*}
    Q =\frac{c^{n}\eta^{n}(\alpha+1)^{m}\alpha^{n-m}(\prod_{i=1}^{m}y_i)^{\eta-1}}{(\prod_{j=m+1}^{n}y_j)^{\alpha \eta+1}}.
    	\end{equation*}
    Therefore, given $\eta$ and $m$ are known, the closed-form solution of $\frac{\partial L(\boldsymbol{y}|\theta, \eta)}{\partial \theta} =0$ can be obtained as:
    \begin{equation}
        \theta(\eta,\boldsymbol{y}) = \frac{(\alpha+1)\sum_{i=1}^{m} y_i^\eta}{(\alpha+1)m-\alpha n}
    \end{equation}
    \par Therefore, we can utilize the search method in section 4.1 to estimate the parameters of an exponentiated exp-Pareto model. 
    
\subsection{Simulations}
To assess the accuracy for the estimates of $\hat{\theta}$ and $\hat{\eta}$ from the search method, a simulation study was done for the chosen sample size $n$, $\theta$ and $\eta$. For each simulation scenario, $\mathit{r} = 2000$ samples were generated from the composite density given in (8). The R package 'mistr'\cite{mistr} was used to generate the random samples from the composite pdf defined in (8). 
\par Table \ref{tab1} to \ref{tab4} present the results from all of the simulation scenarios. $\hat{\eta}_{\text{mean}}$, $\hat{\theta}_{\text{mean}}$  represent the mean of $\hat{\eta}$ and 
$\hat{\theta}$; $\hat{\eta}_{\text{SD}}$ and $\hat{\theta}_{\text{SD}}$ are the standard deviation of $\hat{\eta}$ and $\hat{\theta}$ values, correspondingly. 

\par For all simulation scenarios, we spotted that the average of the estimates of $\theta$ gets closer to the true $\theta$ as the sample size $n$ increases. This is also the case for the average of the estimates of $\eta$. Additionally, for both $\theta$ and $\eta$, the standard deviation of the estimates decreases as the sample size $n$ increases. Thus, the estimates found by this algorithm become more and more accurate as the sample size $n$ increases, which matches the asymptotic property of Maximum Likelihood Estimates (MLE). 
\begin{table}[H] 
\caption{Simulation Results for $\eta = 0.8$ and $\theta = 1$ \label{tab1}}
\newcolumntype{C}{>{\centering\arraybackslash}X}
\begin{tabularx}{\textwidth}{CCCCC}
\toprule
$n$	& $\hat{\eta}_{\text{mean}}$	& $\hat{\theta}_{\text{mean}}$     & $\hat{\eta}_{\text{SD}}$ & $\hat{\theta}_{\text{SD}}$\\\\
\midrule
50		& 0.827615			& 1.046424 & 0.1189729 & 0.3537826\\
100		& 0.811015			& 1.017363 & 0.08019997 & 0.2345774\\
200        &        0.807355       & 1.000019 & 0.05537136 & 0.1534154   \\
\bottomrule
\end{tabularx}
\end{table}
\unskip

\begin{table}[H] 
\caption{Simulation Results for $\eta = 5$ and $\theta = 1$ \label{tab2}}
\newcolumntype{C}{>{\centering\arraybackslash}X}
\begin{tabularx}{\textwidth}{CCCCC}
\toprule
$n$	& $\hat{\eta}_{\text{mean}}$	& $\hat{\theta}_{\text{mean}}$     & $\hat{\eta}_{\text{SD}}$ & $\hat{\theta}_{\text{SD}}$\\\\
\midrule
50		& 5.158100			& 1.046424 & 0.7489789 & 0.3469250\\
100		& 5.076575			& 1.017905 & 0.50643771 & 0.2419570\\
200        & 5.042250              &  1.006531 & 0.35411868 & 0.1578944  \\
\bottomrule
\end{tabularx}
\end{table}
\unskip

\begin{table}[H] 
\caption{Simulation Results for $\eta = 0.8$ and $\theta = 5$ \label{tab3}}
\newcolumntype{C}{>{\centering\arraybackslash}X}
\begin{tabularx}{\textwidth}{CCCCC}
\toprule
$n$	& $\hat{\eta}_{\text{mean}}$	& $\hat{\theta}_{\text{mean}}$     & $\hat{\eta}_{\text{SD}}$ & $\hat{\theta}_{\text{SD}}$\\\\
\midrule
50	& 0.827620			& 5.551189 & 0.1172865 & 2.0430630\\
100		& 0.814515			& 5.233835 & 0.07802185 & 1.2687714\\
200        & 0.807080              & 5.134532 & 0.05448299 & 0.8504233   \\
\bottomrule
\end{tabularx}
\end{table}
\unskip

\begin{table}[H] 
\caption{Simulation Results for $\eta = 5$ and $\theta = 5$ \label{tab4}}
\newcolumntype{C}{>{\centering\arraybackslash}X}
\begin{tabularx}{\textwidth}{CCCCC}
\toprule
$n$	& $\hat{\eta}_{\text{mean}}$	& $\hat{\theta}_{\text{mean}}$     & $\hat{\eta}_{\text{SD}}$ & $\hat{\theta}_{\text{SD}}$\\\\
\midrule
50		& 5.139675			& 5.502811 & 0.7141230& 1.9971169\\
100		& 5.071900  			& 5.259176 & 0.50519868 & 1.2889135\\
200        &  5.046740             &  5.124653 & 0.35052404 & 0.8586674  \\
\bottomrule
\end{tabularx}
\end{table}
\unskip

\section{Real Data Application}
We applied the exponentiated IG-Pareto model and the exponentiated exp-Pareto model to the well-known Danish Fire Insurance Data and Norwegian Fire Insurance Data. We compared these two models with the original one-parameter IG-Pareto and exp-Pareto models. For additional comparison purposes, we included Weibull, Inverse Gamma, Weibull-Inverse Weibull and Weibull-Pareto models.  
\par We used five different Goodness-of-Fit measures to compare the performances of the different models. The brief description of the measures are listed as follows:
\begin{enumerate}

	\item \textbf{NLL:} The Negative Log-Likelihood is defined as follows:
        \begin{center}
		$NLL = -logL(\boldsymbol{\hat{\theta}|y})$
	\end{center}

	\item \textbf{AIC:} The definition of the Akaike Information Criterion \cite{burnham} is as follows:
\begin{center}
	$AIC = -2logL(\boldsymbol{\hat{\theta}|y})+2k$,
\end{center}
where $k$ is the number of free parameters. 

	\item \textbf{BIC:} The Bayesian Information Criterion \cite{burnham}  is defined as follows:  
	\begin{center}
	$BIC = -2logL(\boldsymbol{\hat{\theta}|y})+klog(n)$,
	\end{center}
where $k$ is the number of parameters and $n$ is the sample size of the data set.  
\item \textbf{CAIC:} The Consistent Akaike Information Criterion \cite{caic_1987} is defined as follows:
\begin{center}
$	CAIC = -2logL(\boldsymbol{\hat{\theta}|y})+k(log(n)+1)$,
\end{center}
\item \textbf{AICc:} The corrected Akaike Information Criterion \cite{aicc_1989} is as follows: 
\begin{center}
	$AICc = -2logL(\boldsymbol{\hat{\theta}|y})+\frac{2nk}{(n-k-1)}$,
\end{center}

\end{enumerate}
\par MLEs of the parameters in all the models, NLL, AIC, BIC, AICc and CAIC values were computed with R software. 

\subsection{Danish Fire Insurance Data}

\par
\begin{figure}[!htbp]
	\centering
	\includegraphics[width=0.7\linewidth]{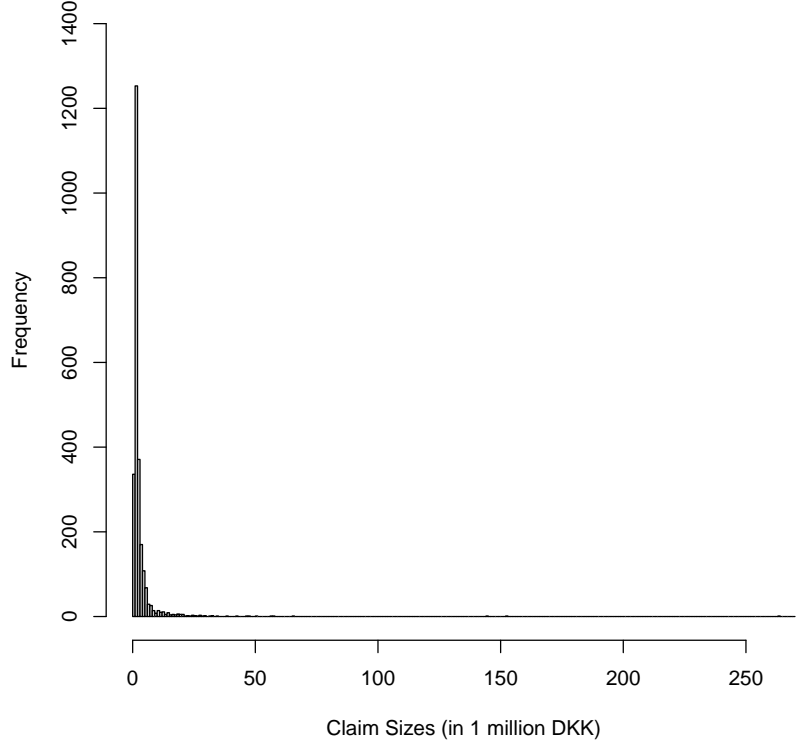}
	\caption{Histogram of Danish Fire Insurance Data Set.}
	\label{fig:danish_hist}
\end{figure}

 To assess the performance of different models, the well-known Danish fire insurance data was used. There are 2492 claims from years 1980 to 1990 in this data set. The data has been scaled in one million Danish Krones (DKK) for analysis purposes. We accessed the data set from the \textit{SMPracticals} package in R software \cite{smpractical}. Figure \ref{fig:danish_hist} is the histogram of this data set. The detailed summary of the data set is provided in Table \ref{tab5}.

\begin{table}[!htbp]
\caption{Summary Statistics for Danish Fire Insurance Data (in 1 million DKKs) \label{tab5}}
\newcolumntype{C}{>{\centering\arraybackslash}X}
\begin{tabularx}{\textwidth}{CCCCCCCC}
\toprule
 Sample Size	& Mean     & SD & Minimum & Q1 & Q2 & Q3 & Maximum \\\\
\midrule
2492 & 3.06 & 7.98 & 0.31 & 1.16 & 1.63 & 2.65 & 263.25 \\
\bottomrule
\end{tabularx}
\end{table}
\unskip

\par
\begin{figure}[!htbp]
	\centering
	\includegraphics[width=0.6\linewidth]{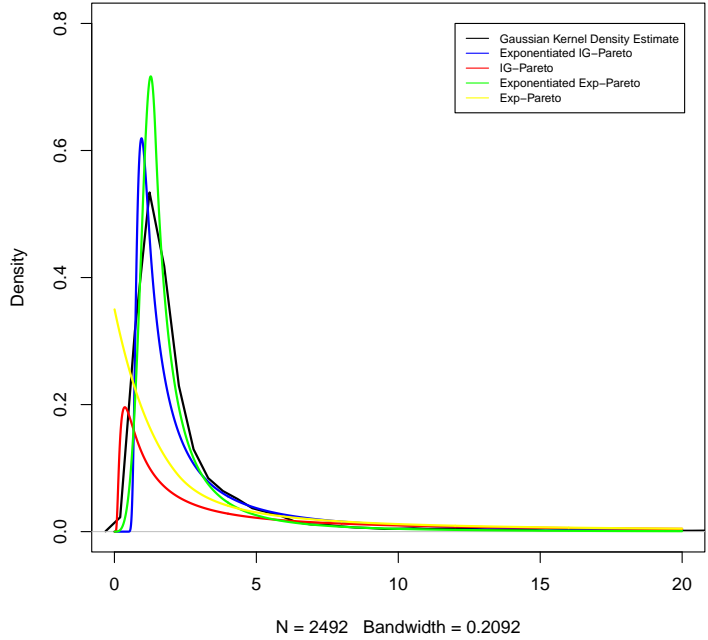}
	\caption{Density Plot of Danish Fire Insurance Data with corresponding exponentiated IG-Pareto, IG-Pareto, exponentiated exp-Pareto and exp-Pareto model fit}
	\label{fig:danish}
\end{figure}

 The model comparisons in terms of all five Goodness-of-Fit measures are presented in Table \ref{tab6}. When fitting to the Danish Fire Insurance Data, the exponentiated IG-Pareto model demonstrated a significantly better performance compared to the original one-parameter IG-Pareto model. Similarly, the exponentiated exp-Pareto model also presented a significantly improvement from the original one-parameter exp-Pareto model in terms of NLL, AIC, BIC, AICc and CAIC. These results are consistent with Figure \ref{fig:danish}.  Figure \ref{fig:danish} shows the comparison of the fitted IG-Pareto model, exponentiated IG-Pareto model, exp-Pareto model, exponentiated exp-Pareto model and the Gaussian kernel density estimate of the Danish Fire Insurance Data. Both exponentiated models provide a good fit to the Danish Fire Insurance Data while the original one-parameter composite models perform poorly. It is also noticeable that the two-parameter exponentiated exp-Pareto model performed better than the common distributions such as Weibull and IG. However, both the Weibull-Inverse Weibull model and the Weibull-Pareto model with the mixing weights still perform slightly better than the proposed exponentiated models. 
\begin{table}[!htbp]
\centering

\caption{Goodness-of-Fit measures of different models of the Danish fire insurance data based on MLEs. \label{tab6}}
	\begin{tabular}{lllllll} 
\hline
			\toprule
		\textbf{Model}                                                                & $\boldsymbol{p}$\textsuperscript{1} & \textbf{NLL} & \textbf{AIC} & \textbf{BIC} & \textbf{AICc} & \textbf{CAIC}  \\
			\midrule
			Weibull & 2                                                                             & 5270.5     & 10545.0~    & 10556.6     & 10545.0      & 10558.6 \\ \hline
		\begin{tabular}[c]{@{}l@{}} IG\end{tabular} & 2                                                                             & 4097.9     & 8199.8~    & 8211.4     & 8199.8     & 8213.4 \\ \hline
			\begin{tabular}[c]{@{}l@{}}Inverse Gamma-Pareto\\(One-Parameter)\end{tabular}	& 1                                                                             & 6983.8     & 13969.6~    & 13975.5~    & 13969.6      & 13976.5   \\
			\hline
		\begin{tabular}[c]{@{}l@{}}Exponentiated\\Inverse Gamma-Pareto\end{tabular}    & 2                                                                             & 4287.7     & 8591.0     & 8590.0     & 8579.4      & 8593.0		\\ \hline
			\begin{tabular}[c]{@{}l@{}}exp-Pareto\\(One-Parameter)\end{tabular} & 1 & 5878.008 & 11758.02 & 11763.84 & 11758.02 & 11764.84  \\
			\hline	\begin{tabular}[c]{@{}l@{}} Exponentiated \\ exp-Pareto\end{tabular} & 2 & 3961.018 & 7926.036 & 7937.678 & 7926.041 & 7939.678\\ \hline
			\begin{tabular}[c]{@{}l@{}} Weibull- \\ Inverse Weibull\end{tabular}\textsuperscript{2}                                                       & 4   & 3820.0      & 7648.0     & 7671.3     & 7648.0      & 7675.3 \\ \hline
	\begin{tabular}[c]{@{}l@{}} Weibull-Pareto\end{tabular}\textsuperscript{2} & 4 & 3823.7      & 7655.4     & 7678.6     & 7655.4      & 7682.5 \\
			\bottomrule
\hline
\end{tabular}\\
		\noindent{ \footnotesize{\textsuperscript{1} $p$ is the number of parameters in the model.}}\\
	\noindent{ \footnotesize{\textsuperscript{2} The composite model has an additional weight parameter $\phi$. \cite{grun2019}}}
\end{table}

\unskip
\pagebreak

\subsection{Norwegian Fire Insurance Dataset: Year 1990}
The Norwegian fire insurance data was utilized in the previous literature to check the performance of different loss models.  The data set contains 9181 claims in 1000s of Norwegian Krones (NKK). We were able to access the data via R package \textit{ReIns} \cite{reins}. Since we do not know if the data was inflation adjusted, we chose only the claims from year 1990 for the analysis. 
\par
\begin{figure}[!htbp]
	\centering
	\includegraphics[width=0.5\linewidth]{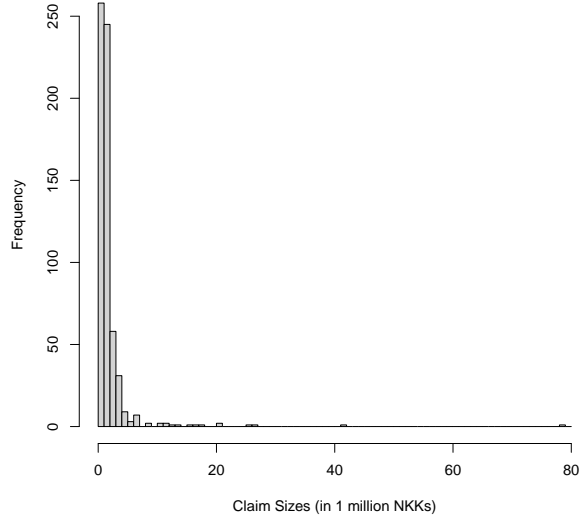}
	\caption{Histogram of Norwegian Fire Insurance Data: Year 1990}
	\label{fig:norway_90}
\end{figure}

 Figure \ref{fig:norway_90} is the histogram of the data set. There are 628 reported claims from year 1990 in the data set. For analysis concern, we scaled the data so that the claim values had a unit of millions of Norwegian Krones (NKK). The detailed summary of the data set is provided in Table \ref{tab7}. 
\begin{table}[!htbp]
\caption{Summary Statistics for Norwegian Fire Insurance Claims During Year 1990 (in 1 million NKKs) \label{tab7}}
\newcolumntype{C}{>{\centering\arraybackslash}X}
\begin{tabularx}{\textwidth}{CCCCCCCC}
\toprule
 Sample Size	& Mean     & SD & Minimum & Q1 & Q2 & Q3 & Maximum \\\\
\midrule
628 & 1.97 & 4.26 & 0.50 & 0.79 & 1.15 & 1.81 & 78.54 \\
\bottomrule
\end{tabularx}
\end{table}
\unskip

\begin{figure}[!htbp]
	\centering
	\includegraphics[width=0.6\linewidth]{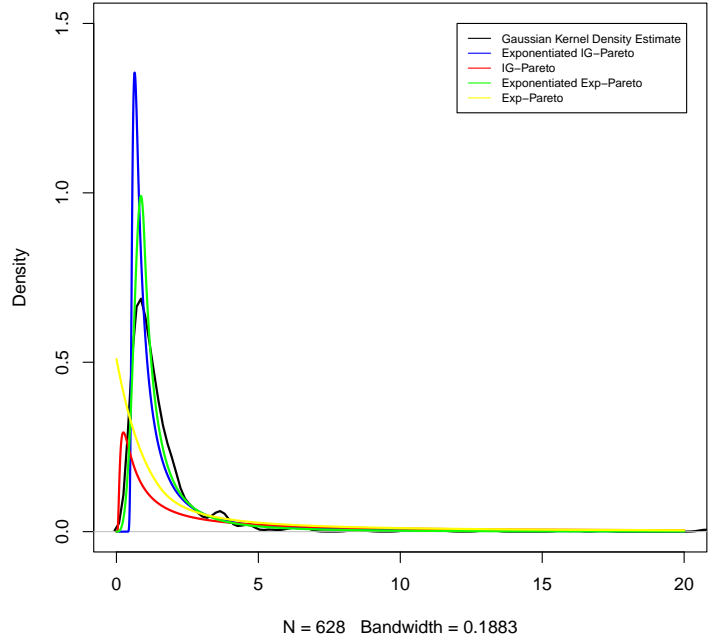}
	\caption{Density Plot of Norwegian Fire Insurance Data from year 1990 with corresponding exponentiated IG-Pareto, IG-Pareto, exponentiated exp-Pareto and exp-Pareto model fit}
	\label{fig:norway_density}
\end{figure}

Table \ref{tab8} summarizes the comparisons of different models when fitting to this data set. Similar to the results that we obtained for Danish Fire Insurance Data, the two proposed exponentiated models significantly overperform the original one-parameter composite models in terms of all Goodness-of-Fit Measures. This is also illustrated visually in Figure \ref{fig:norway_density}. The two exponentiated models provide closer fits the Gaussian kernel density estimate of the data set compared to the original one-parameter composite models. Both of the two proposed two-parameter exponentiated models perform better than Weibull and IG model. In addition, among all eight models we chose for real data application, the exponentiated Inverse Gamma-Pareto model demonstrates the best performance in terms of BIC and CAIC. 
\begin{table}[!htbp]
\centering

\caption{Goodness-of-Fit measures of different models of the Norwegian fire insurance data from year 1990 based on MLEs. \label{tab8}}
\begin{tabular}{lllllll} 
\hline
		\textbf{Model}                                                                & $\boldsymbol{p}$\textsuperscript{1}& \textbf{NLL} & \textbf{AIC} & \textbf{BIC} & \textbf{AICc} & \textbf{CAIC}  \\
			\midrule
			Weibull & 2   & 1054.423 & 2112.846 & 2121.731 & 2112.865 & 2123.731                                                                           \\ \hline
		\begin{tabular}[c]{@{}l@{}} IG\end{tabular} & 2     & 773.8578 & 1551.716 & 1560.601 & 1551.735 & 1562.601                                                                         \\ \hline
			\begin{tabular}[c]{@{}l@{}}Inverse Gamma-Pareto\\(One-Parameter)\end{tabular}	& 1   & 1503.705 & 3009.410 & 2013.853 & 3009.416 & 3014.853                                                                             \\
			\hline
		\begin{tabular}[c]{@{}l@{}}Exponentiated\\Inverse Gamma-Pareto\end{tabular}    & 2                                                                             & 755.5741     & 1515.148     & 1524.033     & 1515.167     & 1526.033		 \\ \hline
			\begin{tabular}[c]{@{}l@{}}exp-Pareto\\(One-Parameter)\end{tabular} & 1 & 1225.63 & 2453.260 & 2457.703 & 2453.266 & 2458.703  \\
			\hline	\begin{tabular}[c]{@{}l@{}} Exponentiated \\ exp-Pareto\end{tabular} & 2 & 772.2026 & 1548.405 & 1557.290 & 1548.424 & 1559.290\\ \hline
			\begin{tabular}[c]{@{}l@{}} Weibull- \\ Inverse Weibull\end{tabular}\textsuperscript{2}                                                       & 4 & 750.9702 & 1509.940 & 1527.711 & 1510.005 & 1531.711   \\ \hline
	\begin{tabular}[c]{@{}l@{}} Weibull-Pareto\end{tabular}\textsuperscript{2} & 4 & 1077.078 & 2162.156 & 2179.926 & 2162.220 & 2183.926  \\
			\bottomrule
\hline
\end{tabular}\\
		\noindent{ \footnotesize{\textsuperscript{1} $p$ is the number of parameters in the model.}}\\
	\noindent{\footnotesize{\textsuperscript{2} The composite model has an additional weight parameter $\phi$. \cite{grun2019}}}
\end{table}
\unskip

\section{Discussion}
\par In this paper, we proposed a generalized family of exponentiated composite distributions. The motivation of proposing such family is to improve the flexibility of the original composite distribution by introducing an exponent parameter. Similar to how the Weibull distribution was developed based on exponential distribution, we introduce the exponent parameter by exponentiating a general random variable associated with a composite distribution defined in Section 2.1. We proved that an exponentiated composite distribution is still a composite distribution and we derived some mathematical properties of this new family of distribution, including the raw moments and the limited moments. The two-parameter exponentiated IG-Pareto and exponentiated exp-Pareto model were discussed as the special models within this family. We also provided a method to find the estimates of the parameters in an exponentiated composite model. The simulations in section 4 showed that the method has the ability to identify the true parameters for an exponentiated exp-Pareto model. We assess the performances of the two-parameter exponentiated Inverse Gamma-Pareto and exponentiated exp-Pareto model with two insurance data sets and compared their performances to some existing models in the past literature. For the Norwegian Fire Insurance Data during year 1990, the exponentiated IG-Pareto model demonstrates the best performance among all of the models we chose in terms of BIC and CAIC. We foresee this family of exponentiated models will significantly increase the abundance of the composite models. Since we also proved the exponentiating composite models are still composite models, this family can be extended to the exponentiated composite models with the mixing weights. 
\bibliography{ref}
\bibliographystyle{unsrt}

\end{document}